\documentstyle[12pt]{article}
\begin{document}
\begin{center}
{\bf On the second-neighbor correlator in 1D XXX quantum antiferromagnetic
 spin chain}\\
\vspace*{1cm}
{\sc J. Dittrich}\\
{\it Nuclear Physics Institute, Academy of Sciences of the Czech Republic,
CZ-250 68 \v Re\v z, Czech Republic}
\footnote{Also member of the Doppler Institute
of Mathematical Physics, Faculty of Nuclear Sciences and Physical Engineering,
Czech Technical University, Prague}
\\
and\\
{\sc V.I. Inozemtsev}\\
{\it BLTP, JINR, 141980 Dubna, Russia}\\
\vspace*{2cm}
Abstract\\
\end{center}

We have calculated the energy per site for the ground state of
 antiferromagnetic quantum
spin chain with variable range exchange $h(j-k)\propto \sinh^2 a
\sinh^{-2}a(j-k)$ in the framework of the asymptotic Bethe ansatz.
By expanding it in powers of $e^{-2a}$, we have confirmed the value of the
second-neighbor correlator for the model with nearest-neighbor exchange
obtained earlier in the atomic limit of the Hubbard chain.
\newpage

At present, there are two main approaches in investigations of spin
correlations in the ground state of the quantum Heisenberg antiferromagnet.
The first one is connected with analysis of the representations of quantum
groups [1] and gives some universal prescription of calculation of various
correlations via multiple contour integrals. Unfortunately, till now there
are no possibilities to reduce these multiple integrals to known
transcendents. The second approach consists in finding correlators in the form
of Fredholm determinants [2]. However, within this approach there is still
 no way
to analytic evaluation of these Fredholm determinants. Asymptotics of spin
correlations at large distances between spins has been obtained in the
framework of conformal field theory [3], but this theory cannot give the 
values of correlators at finite distances.

The first explicit calculation of nearest-neighbor spin correlator
$<\vec \sigma_{j}\vec \sigma_{j+1}>$ in the thermodynamic limit has been
made by Hulthen [4] with the use of the Bethe ansatz [5]. Since this
correlator coincides with the energy per site in the vacuum state of the 
model, the problem is equivalent to calculation of this energy for finite
number of lattice sites $N$ and taking thermodynamic limit $N\to\infty$.
The second-neighbor correlator $<\vec \sigma_{j}\vec \sigma_{j+2}>$ has been
calculated by Takahashi [6] with the use of perturbation theory in the atomic
limit of the half-filled Hubbard model. This result has never been confirmed
by other scheme of calculations.

In this letter, we perform the explicit calculation of $<\vec \sigma_{j}
\vec \sigma_{j+2}>$ by using the way which differs substantially from the way
of ref.[6]. Namely, we start from the consideration of the integrable
spin chain with variable range exchange and the Hamiltonian [7]
$${\cal H}={1\over 2}\sum_{j\neq k}{{\sinh^2{a}}\over{\sinh^2 a(j-k)}}
{{\vec \sigma_{j}\vec \sigma_{k}-1}\over 2}.
\eqno(1)$$
In the nearest-neighbor limit $a\to\infty$ one can decompose (1) into
a series
$${\cal H}= {1\over 2}\sum_{j}(\vec \sigma_{j}\vec \sigma_{j+1}
-1)+{1\over 2}e^{-2a}\sum_{j}(\vec \sigma_{j}\vec \sigma_{j+2}-1)
+o(e^{-2a}).\eqno(2)$$
Hence one can write the ground-state energy per site as
$$e={1\over 2}<\vec \sigma_{j}\vec \sigma_{j+1}-1>
+{1\over 2}e^{-2a}<\vec \sigma_{j}\vec \sigma_{j+2}-1> +o(e^{-2a}),\eqno(3)
$$
where $<$ $>$ means average on the vacuum state of the Hamiltonian (2).
Fortunately, in the first order approximation (3) one can replace this state
to the vacuum state of nonperturbed Hamiltonian with the interaction
of nearest-neighbor spins, ${\cal H}_{0}={1\over 2}\sum_{j}
(\vec \sigma_{j}\vec \sigma_{j+1}-1)$.

Let us now calculate the ground-state energy per site (3) of the model
with the use of the asymptotic Bethe ansatz (ABA) [8]. The wave functions of 
the states with $M$ down spins can be calculated exactly [7]. Their
asymptotic expression in the region $n_{1}\ll n_{2}...\ll n_{M}$
has the form similar to Bethe one,
$$\psi(n_{1},..n_{M})\propto\sum_{P\in \pi_{M}}
\exp\left(i\sum_{\alpha=1}^M k_{P\alpha}n_{\alpha}\right)
\exp\left({i\over 2}\sum_{\alpha<\beta}^M \chi(k_{P\alpha},k_{P\beta})\right
),\eqno(4) $$
where the first sum is taken over all permutations from the group $\pi_{M}$,
$\{k_{\alpha}\}$ is the set of pseudomomenta and $\chi(k_{\alpha},k_{\beta})$
is the two-magnon phase shift defined by the relations
$$\cot{{\chi(k_{\alpha},k_{\beta})}\over 2}=\varphi(k_{\alpha})-
\varphi(k_{\beta}),\eqno(5)$$
$$\varphi(k)={k\over{2\pi i a}}\zeta_{1}\left({{i\pi}\over{2a}}\right)-
{1\over {2ia}}\zeta_{1}\left({{ik}\over {2a}}\right),\eqno(6)$$
where $\zeta_{1}$ is the  zeta Weierstrass function defined on the torus
${\bf T}_{1}={\bf C}/({\bf Z}+{{i\pi}\over a}{\bf Z})$.
To consider the chains of finite length $N$ in the thermodynamic limit
$N\to\infty$, we adopt the main hypothesis of ABA, i.e. imposing periodic
boundary conditions on the asymptotic form of the wave function (4) (it is
worth noting here that the exact solution of the problem of finite chains in
the model should be based on the treatment of the Hamiltonian with the
 exchange
given by the Weierstrass $\wp$ function instead of (1), and is not available
till now except the cases $M=2,3$). 
Taking $\psi(n_{2},..n_{M},n_{1}+N)=\psi(n_{1},..n_{M})$ and calculating 
the both  sides
with the use of the formula (4) results in the ABA equations
$$\exp(ik_{\alpha}N)=\exp\left(i\sum_{\beta\neq\alpha}^{M}\chi(k_{\alpha},
k_{\beta})\right),\qquad \alpha=1,..M.\eqno(7)$$
The energy of corresponding configuration is given by
$$E_{M}=\sum_{\alpha=1}^{M}\sum_{n\neq 0}{{\sinh^2 a}\over{\sinh^2 an}}
(\cos(k_{\alpha}n)-1).\eqno (8)$$
We are interested in the antiferromagnetic vacuum of the model and should
take $N$ even, $M=N/2$.
Taking logarithms of both sides of (7) and choosing the proper branches
 [9], 
one arrives at
$${{Q_{\alpha}}\over N}={{\pi-k_{\alpha}}\over{2\pi}}-
{1\over{\pi N}}\sum_{\beta\neq\alpha}^M \arctan[\varphi(k_{\alpha})
-\varphi(k_{\beta})].\eqno(9)$$
We adopt usual hypothesis about the distribution of the
(half)integers $\{Q\}$ for antiferromagnetic vacuum state. Namely, it
will be assumed that these numbers form uniform string from $-Q_{max}$ to
$Q_{max}$, $Q_{max}=N/4-1/2$ without holes. We introduce the rapidity
variable $\lambda$ by the relation $\lambda=\varphi(k)$ and the function
$\mu(\lambda)$ via the relation $\pi -k=\mu(\lambda)$. The ABA equations
(9) now can be written as
$$Q_{\alpha}/N=Z(\lambda_{\alpha}),\eqno(10)$$
where
$$Z(\lambda)=(2\pi)^{-1}\mu(\lambda)-{1\over{\pi N}}\sum_{\beta=1}\arctan
(\lambda-\lambda_{\beta}).$$
Following Hulthen [4], let us go to continuous variable $x=Q_{\alpha}/N$
in the limit $N\to\infty$ and introduce the root density $\sigma_{N}(\lambda)
$ by the relation $\sigma_{N}(\lambda)=dx/d\lambda$. Differentiating both
sides of (10) with respect to $\lambda$, one arrives at the Hulthen-like
equation in the limit $N\to\infty$
$$\sigma_{\infty}(\lambda)=(2\pi)^{-1}\mu'(\lambda)-
\int_{-\infty}^{\infty}A(\lambda-\lambda')\sigma_{\infty}(\lambda')d\lambda'
,\eqno(11)$$
where $A(\lambda)=[\pi(1+\lambda^2)]^{-1}$. The energy per site can be 
written from (8) as
$$e_{\infty}=\lim_{N\to\infty}N^{-1}E_{N/2}=\int_{-\infty}^{\infty}
\varepsilon(k(\lambda))\sigma_{\infty}(\lambda)d\lambda,\eqno(12)$$
where
$$\varepsilon(k(\lambda))=2\sinh^2 a\sum_{n=1}^{\infty}
{{\cos nk(\lambda)-1}\over{\sinh^2 an}}.\eqno(13)$$
The solution to (11) can be found via Fourier transform,
$$\sigma_{\infty}(\lambda)=(2\pi)^{-2}\int_{-\infty}^{\infty}
{{e^{i\lambda p}dp}\over{1+e^{-\vert p\vert}}}\int_{-\infty}^{\infty}
\mu'(\tau)e^{-ip\tau}d\tau.$$ 
Substituting it into (12) yields
$$e_{\infty}=(2\pi)^{-2}\int_{-\infty}^{\infty}d\lambda\varepsilon(k(\lambda))
\int_{-\infty}^{\infty}dp{{e^{ip\lambda}}\over {1+e^{-\vert p\vert}}}
\int_{-\infty}^{\infty}\mu'(\tau)e^{-ip\tau}d\tau.$$
 Choosing variables as $\lambda=\varphi(k),\mu'(\tau)d\tau=-dk'$
and changing the order of integration (it is allowed since the integral
over $\tau$ vanishes sufficiently fast as $\vert p\vert\to\infty$),
 one arrives at the main formula
$$e_{\infty}=-(2\pi)^{-2}\int_{-\infty}^{\infty}{{dp}\over{1+e^{-\vert p\vert
}}}\int_{0}^{2\pi}dk \varepsilon(k)\varphi'(k)e^{ip\varphi(k)}
\int_{0}^{2\pi}dk'e^{-ip\varphi(k')},\eqno (14)$$
where the functions $\varepsilon (k)$ and $\varphi(k)$ are determined by
(13) and (6). The integrals in (14) cannot be calculated analytically.
However, one can see from (6) and (13) that for large $a$
$$\varphi(k)={1\over 2}\cot{k\over 2} + 2e^{-2a}\sin k+o(e^{-2a}),
\eqno(15)$$
$$\varepsilon (k) =2(\cos k-1)+2e^{-2a}(\cos 2k-1)+o(e^{-2a}).\eqno(16)$$
Substituting (15) and (16) into (14), one can calculate inner integrals
up to the order of $e^{-2a}$,
$$\int_{0}^{2\pi}dk'e^{-ip\varphi(k')}=2\pi(1-p^2 e^{-2a})e^{-\vert p/2\vert}
+o(e^{-2a}),$$
$$\int_{0}^{2\pi}dk \varepsilon(k)\varphi'(k)e^{ip\varphi(k)}
=\pi(2+e^{-2a}(8-p^2))e^{-\vert p/2\vert} +o(e^{-2a}).$$
Now, taking into account the formula $\int_{0}^{\infty}{{p^2 dp}\over
{1+e^p}}=3/2\zeta(3)$, where $\zeta$ is the Riemann zeta function,
the two leading terms of $e_{\infty}$ are calculated. The comparison of
 (3) and (14) then gives
$$<\vec \sigma_{j}\vec \sigma_{j+2}>=1 -16 \ln 2 +9\zeta(3),\eqno(17)$$
which coincides exactly with the expression given by Takahashi [6].

To conclude, we would like to point out that our approach to the calculation
of the correlator is based on the hypothesis of the validity of the ABA
and uniform distribution of the numbers $Q_{\alpha}$ for the antiferromagnetic
vacuum of the model (1). The approach of ref.[6] refers to the perturbation 
theory in the atomic limit of half-filled Hubbard model and the integral
representation of its ground-state energy in the thermodynamic limit
 which is derived under the
 assumption on uniform distribution of integers in the Bethe ansatz equations
for the model. Hence both results are not absolutely
 rigorous from mathematical point
of view, and their coincidence supports the validity of both approaches.
Unfortunately, neither our method nor the method of Takahashi [6] can
be used for the calculations of higher correlators. It seems that these
calculations will be possible if one will find the method of reduction of
 multiple integrals in general formulae of [4]. Till now, there is no recipe
of doing this even for the second-neighbor correlator (17).\\

\noindent
The work is partly supported by the GACR grant No. 202/96/0218. 
\newpage
{\bf References}

\begin{enumerate}
\item
Jimbo M and Miwa T 1994 {\it Algebraic Analysis of Solvable Lattice Models}
 RIMS-981
\item
Korepin V E, Izergin A G, Essler F H L and Uglov D B 1994
{\it Phys.Lett.} {\bf A190} 182 
\item
Affleck J 1986 {\it Nucl.Phys.} {\bf B265} 409 
\item
Hulthen L 1938 {\it Ark.Mat.Astron.Fys.} {\bf A26} (11)
\item
Bethe H A 1931 {\it Z.Phys.} {\bf 71} 205
\item
Takahashi M 1977 {\it J.Phys} {\bf C10} 1289 
\item 
Inozemtsev V I 1992 {\it Commun.Math.Phys.} {\bf 148} 359
\item
Sutherland B 1979 {\it Rocky Mountain J.Math.} {\bf 8} 413
\item
Hamer C  1986 {\it J.Phys.A:Math.Gen.} {\bf 19} 3335 
\end{enumerate}

\end{document}